\documentclass[12pt]{article}
\usepackage{times}             
\usepackage[latin1]{inputenc}  
\usepackage{graphicx}          
\begin{document}
\begin{center}
{\large\bf 
Impurity modes and effect of clustering in diluted semiconductor alloys}
\\
\medskip
\underline{Andrei Postnikov}, Olivier Pag\`es, Ayoub Nassour and Joseph Hugel
\\
\emph{LPMD, Paul Verlaine University -- Metz, 1 Bd Arago, F-57078 Metz, France}
\\
postnikov@univ-metz.fr
\end{center}
\medskip
The variation of TO zone-center vibration spectra with concentration
in mixed zincblende-type semiconductors can be understood within
a paradigm of unified ``one bond -- two modes'' approach,
which has been recently outlined as a rather general concept,$^1$
and emerges from a number of previous experimental and
theoretical studies.$^{2,3}$ The crucial issue is 
that the vibration frequency, associated with a certain cation-anion bond,
depends on the length of the latter,
and the bond length, in its turn, depends not
only on the average alloy concentration, but on local variations of it.
In an (A,B)C substitutional alloy,
the A--C bond length differ in A-rich and A-poor regions,
yielding a splitting of the A--C vibration frequency.
Such splittings can be measured and reproduced
in first-principles calculations. 

\begin{figure}[b]
\noindent
\parbox[b]{11.4cm}{\includegraphics[width=11.0cm]{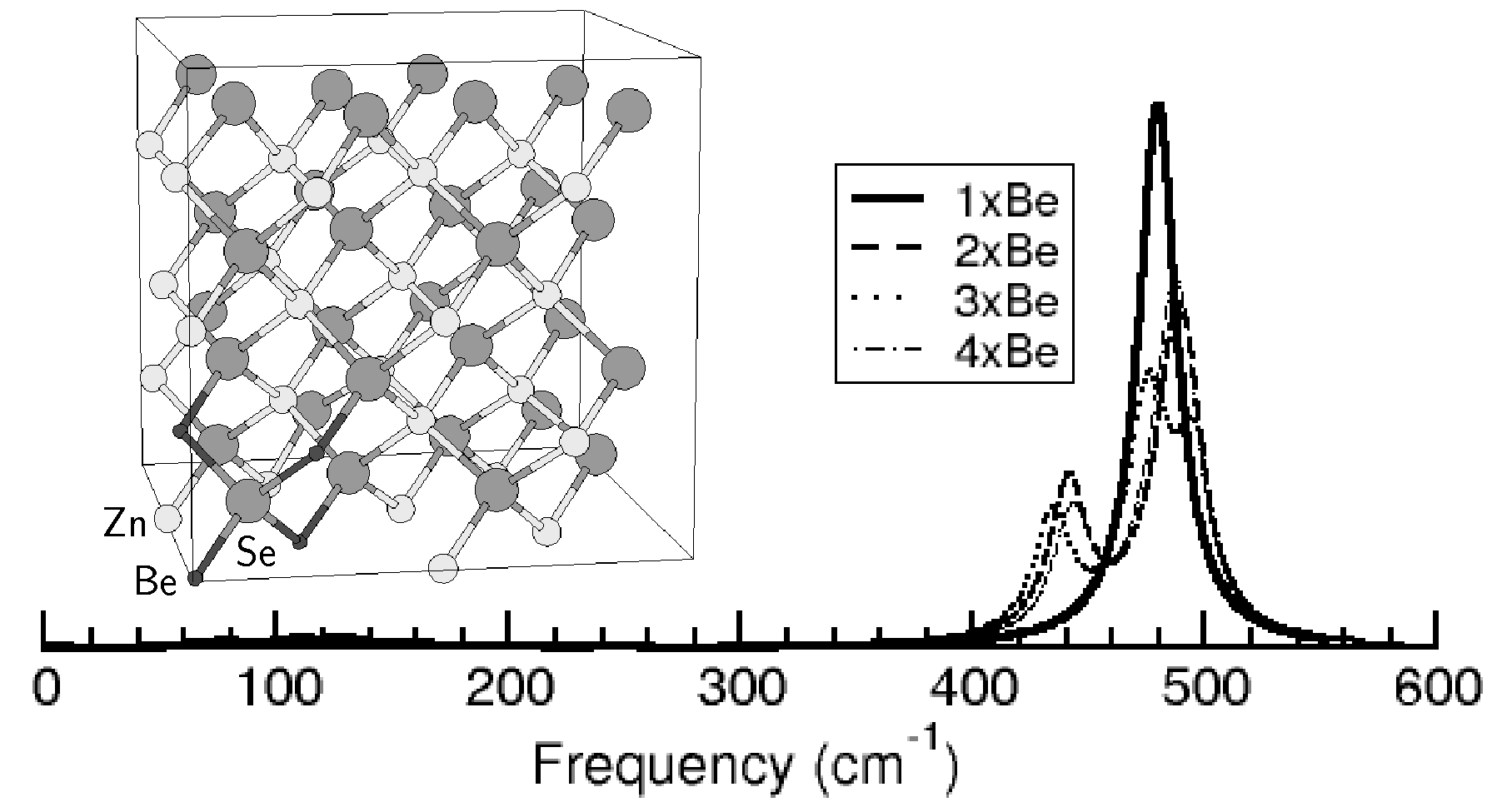}}
\parbox[b]{4.4cm}{\footnotesize
Fig.~1. Phonon density of states of Be, after first-principles
calculations for Be$_n$Zn$_{32-n}$Se$_{32}$ supercells,
with 1,..4 Be cations neighbouring the same Se anion.
The supercell for $n$=4 is shown in the inset. \\*[10mm]
}
\end{figure}

An analysis of vibration spectra helps to get an insight
into the structural short-range (clustering)
and long-range (formation of extended chains of certain cation-anion
pairs and other structural motives at the mesoscopic scale) tendencies.
For this however, one needs first-principles benchmark calculations 
for representative 
model systems (see, e.g., Ref.~4 for the ZnSe--BeSe alloys).
The simplest yet important result from first-principles calculations is
a prediction of how the impurity phonon mode
evolves as isolated (distant) impurities get clustered.

In the present contribution, we outline the results of first-principles
calculations of phonon frequencies and vibration patterns,
in the dilution limits of several mixed semiconductor alloys,
Be$_x$Zn$_{1-x}$Se, Ga$_x$In$_{1-x}$As and 
Ga$_x$In$_{1-x}$P. The calculations have been done
by the {\sc Siesta} method$^5$
for cubic 64-atom supercells, with one or two cation atoms substituted by
impurity species, that corresponds to impurity contents of 3\% and 6\%,
respectively.
The initial unconstrained structure relaxation for each supercell chosen
was followed by a calculation of phonons by finite displacement
technique. Each of the atoms in the supercell were subject to
6 consecutive small cartesian displacements, and the forces
induced on all atoms in the supercell resulted in corresponding force
constants.

\begin{figure}[b!]
\noindent
\parbox[b]{11.0cm}{\includegraphics[width=10.5cm]{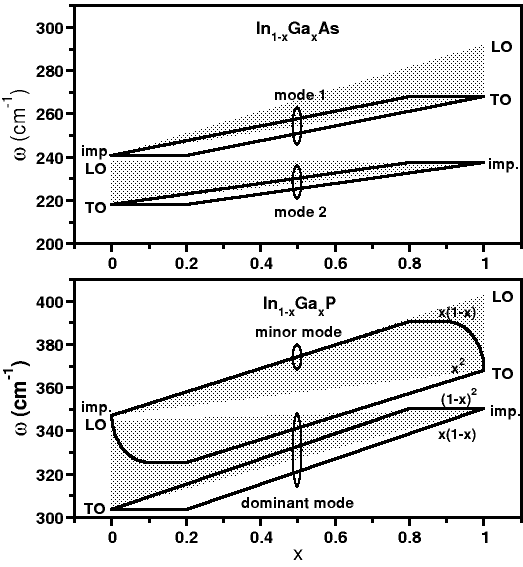}}
\parbox[b]{4.5cm}{\footnotesize%
Fig.~2. Simplified 1-bond$\rightarrow$2-mode TO (thick lines) percolation
schemes of (InGa)As and (InGa)P. The generic fraction
of bonds corresponding to each TO branch is indicated
in the bottom panel. The bond-related optical bands simply obtained by
linear convergence of the parent TO and LO frequencies onto the
related impurity frequencies are shown as shaded areas. \\*[3mm]}
\end{figure}

The analysis of results is the simplest for Be$_x$Zn$_{1-x}$Se,
a system with large contrast in masses and elastic properties between
its parent compounds, that leads to a big separation between
Zn-related and Be-related phonon modes (see Ref.~4 for details).
Fig.~1 shows the phonon density of states for Be, for the cases
of single impurity and impurity pair (both Be atoms being neighbours
to the same Se). Moreover, the cases of three and four Be atoms
grouping around the same Se are considered for this system
(the Be$_4$Zn$_{28}$Se$_{32}$ supercell is shown in the inset to Fig.~1).
One clearly sees that the single vibration peak of an isolated
impurity (a nearly triply degenerate mode of Be in tetrahedral Se$_4$
cage) splits into two peaks, separated by $\sim$40~cm$^{-1}$, 
as another Be impurity is introduced.
The origin for this splitting is a diversification of Be--Se bond lengths,
as two Be share the same Se atom. Namely, the bonds to outer Se atoms
of the Be--Se--Be cluster can be much more efficiently shortened
in the course of structure relaxation than the bonds to the central Se.
Shorter bonds imply larger force constants and hence higher vibration
frequency (again, see Ref.~4 for a more detailed discussion).
It is noteworthy that an addition of further Be atoms as neighbors
to the central Se brings in some additional
structure, but leaves in place the initial major splitting into
two big groups of peaks, those characterizing a ``Be-poor''
environment (at $\sim$480~cm$^{-1}$) and ``Be-rich'' environment
(at $\sim$440~cm$^{-1}$).

In the following we compare this initial splitting of single-impurity
mode into a doublet of interacting-impurities modes with the
experimentally observed onset of the ``1-bond$\rightarrow$2-mode''
behaviour, discussed previously.$^{1-4}$
Notably, the impurity mode at $\sim$440~cm$^{-1}$ in the 
$x{\rightarrow}0$ limit of Be$_x$Zn$_{1-x}$Se develops into
two branches, at $\sim$450~cm$^{-1}$ and $\sim$410~cm$^{-1}$,
at small values of $x$ -- see Fig.~1 of Ref.~6. 
It should be noted (and was already discussed
in Ref.~4) that the absolute values frequencies are shifted upwards
in our calculation, with respect to their experiment values, due to
a slight overbinding caused by the local density approximation
to the exchange-correlation. However, the frequency difference between
the single-impurity and double-impurity modes is fairly reproduced.

The general behaviour of phonon modes depending on the Be concentration, 
``streamlining'' the experimental details from Fig.~~1 of Ref.~6,
is shown and discussed in Ref.~7. It was argued$^1$
that the similar schema holds for other semiconductor alloys,
even if the position of branches and their splittings may differ.
Fig.~2 reproduces a part of experimental findings schematically
presented in Fig.~4 of Ref.~1.
Below we discuss the \emph{ab initio} predictions of the
single impurity vs. impurity pair -related phonon behaviour
for (Ga,In)As and (Ga,In)P alloys.

For the identification of zone-center phonon modes
it is convenient to refer to ${\bf q}$-projected phonon density
of states (PhDOS), which can be introduced as \\*[-1mm]
$$
I_{\aleph}(\omega,{\bf q}) = \sum_i \Bigl| \sum_{\alpha \in \aleph}
A^{\alpha}_i(\omega)\exp({\bf q}{\bf R}_{\alpha})\Bigr|^2
$$

\vspace*{-1mm}
\noindent
where $A^{\alpha}_i(\omega)$ is phonon eigenvector for atom
$\alpha$ at frequency $\omega$, and $\aleph$ -- an arbitrarily
chosen group of atoms (say, those of a given chemical species).
The ${\bf q}$=0 -projected PhDOS of a host atom selects the zone-center
vibrations which contribute to the Raman spectra, like e.g.
a TO-mode of bulk InAs at $\sim$220~cm$^{-1}$ is prominent
in the PhDOS of In in Fig.~3.
For Ga, which is a single impurity, the ${\bf q}$-projected PhDOS 
coincides with the full vibrational density of states and spans
the whole interval of frequencies, from 0 to 250~cm$^{-1}$.
Yet, not all of these modes are zone-center-like,
if one takes into account the vibration of the As sublattice.
A direct inspection of different vibration patterns
reveals, as genuine zone-center vibration modes, a triplet at 246~cm$^{-1}$.
In the ${\bf q}$=0 -projected PhDOS of In (Fig.~3) and of As 
(not shown, as it is practically identical with that of In), a split-off
peak on the high-frequency side of the main TO line does also reveal
the presence of this impurity mode, which is an 
in-phase vibration of Ga impurity with the In sublattice,
in counter-phase to the As sublattice.

Turning to the analysis of impurity pair, we find several
modes, which correspond to in-phase vibration of Ga atoms
with the In sublattice and in counter-phase with the As sublattice. 
These modes span the range of 248--251~cm$^{-1}$.

On the opposite end of the concentration scale, the vibration
of isolated In in GaAs, as a heavy impurity, mixes up with 
a ``continuum'' of bulk modes. Still, the major impurity modes 
can be identified near 238~cm$^{-1}$. The impurity pairs vibrate
in phase between themselves and with the cation sublattice
in the region 235--241~cm$^{-1}$, again in agreement with
the measured\linebreak
\\
\parbox[b]{7.5cm}{%
\vspace*{-3.4mm}
vibration zone-center frequencies shown in Fig.~2.
The first-principles calculations yield therefore, 
on both low-concentration sides of the concentration scale of (Ga,In)As,
the zero splitting from single-impurity frequency to the two-mode regime.
This is consistent with experimental observations summarized
in Fig.~2. \\*[13.4mm]  
{\footnotesize%
Fig.~3. ${\bf q}$=0 -projected phonon density of states \linebreak
of Ga and In in Ga$_1$In$_{31}$As$_{32}$ supercell. \\*[3mm]
}}
\hspace*{5mm}
\parbox[b]{ 8.0cm}{
\vspace*{2mm}
\includegraphics[width=8.0cm]{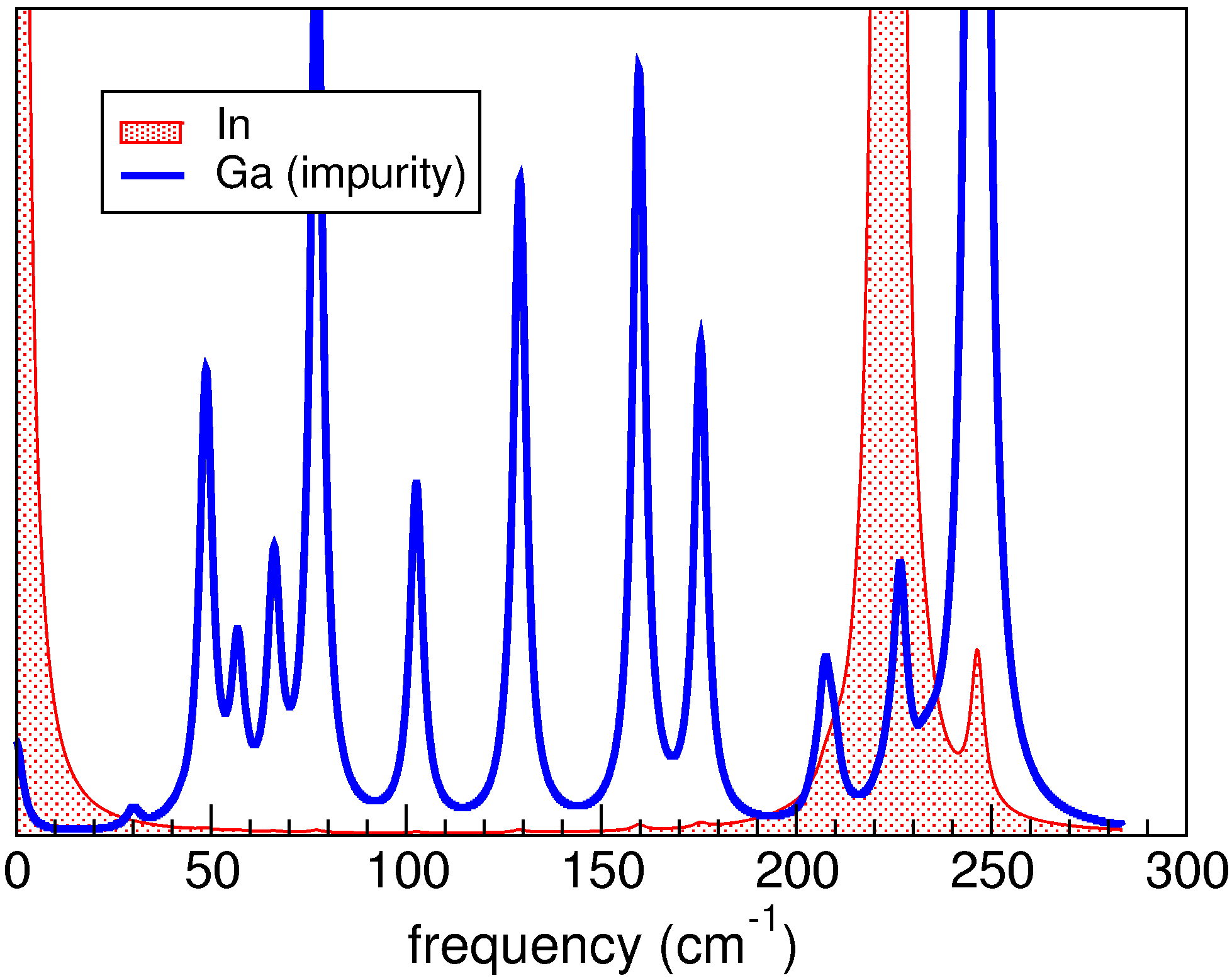}
}

\begin{figure}[h!]
\noindent
\parbox[b]{12.0cm}{\includegraphics[width=11.6cm]{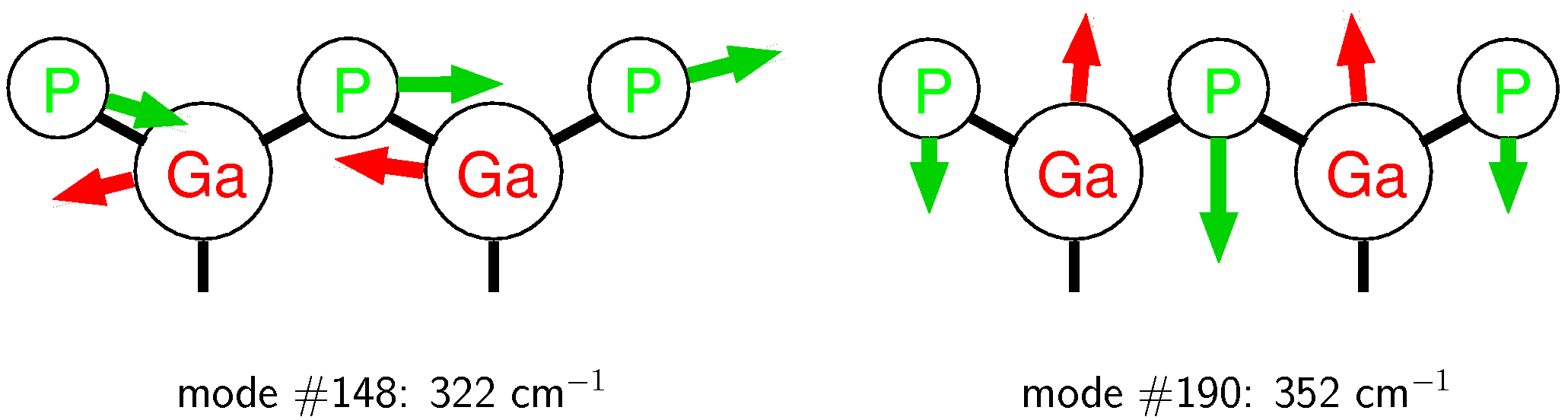}}
\hspace*{2mm}
\parbox[b]{3.6cm}{\footnotesize%
\sloppypar
Fig.~4. Two in-phase vib\-ra\-tions of Ga impurity pair with strong
zone-center contribution in the vibrations of the anion sublattice,
according to calculations in the Ga$_2$In$_{30}$P$_{32}$ supercell.
}
\end{figure}

For (Ga,In)P the situation is different. In the experiment,
the two-mode regime which sets on under the Ga doping is characterized
by a quite large splitting (of $\sim$20~cm$^{-1}$) between the two 
Ga--P modes. 
Indeed we could recover this splitting in the calculation.
Fig.~4 shows the vibration patterns of two maximally split in-phase
double-impurity modes. The softest of these modes (322~cm$^{-1}$) is
a ``longitudinal'' displacement of two Ga atoms against the
P atom in between, making one Ga--P bond longer
at the expense of the other. The hardest mode (352~cm$^{-1}$)
involves a different ``two Ga against P'' displacement, which lies 
in the Ga--P--Ga plane
as in the first case, but occurs in the perpendicular direction,
bending the Ga--P--Ga angle. This vibration is slightly harder
than the single-Ga impurity mode whose calculated frequency is
350~cm$^{-1}$. Taken together, the arrangement of calculated
vibration modes agrees well with the experimental observations
summarized in Fig.~2 (up to a small systematic blue shift of calculated
frequencies with respect to experiment, discussed above).

On the Ga-rich side of the schema in the lower panel of Fig.~2,
no noticeable splitting was detected in experiment, and again
this is consistent with the calculation: single-In impurity modes
at 360~cm$^{-1}$ develop, on addition of a second In impurity,
into a bunch of two-impurity modes, which are confined
in a quite narrow range, 357 to 362~cm$^{-1}$.

Summarizing, we have shown, on discussing impurity limits
in three semiconductor alloys, how the important parameters
of the concentration--frequency diagram for mixed semiconductors
discussed in Ref.~1,
the single impurity frequencies and their initial splitting
on the onset of one-bond$\rightarrow$two-mode behaviour
can be extracted from first-principles calculations.

We acknowledge the access to the calculation resources
at the CINES in Montpellier (projet N$^{\circ}$~pli2623).
    
\bigskip
\noindent
{\footnotesize
\begin{tabular}{r@{.\hspace*{3mm}}p{15.0cm}}
1 & O.~Pag\`es, M.~Kassem, A.~Chafi, A.~Nassour, S.~Doyen and A.~V.~Postnikov,
\newline
    {\tt http://arxiv.org/abs/0709.0930} \\
2 & O. Pag\`es, M. Ajjoun, D. Bormann, C. Chauvet, E. Tourni\'e 
    and J. P. Faurie,
    Phys.~Rev.~B {\bf 65}, 035213 (2002). \\
3 & O. Pag\`es, T. Tite, K. Kim, P.A. Graf, O. Maksimov and M.C. Tamargo,
    \newline
    J.Phys.: Condens.Matter {\bf 18}, 577 (2006). \\
%
4 & A.V. Postnikov, O.~Pag\`es and J.~Hugel,
    Phys.~Rev.~B {\bf 71}, 115206 (2005). \\
5 & {\tt http:/$\!\!$/www.uam.es/siesta};
    P. Ordejon, E. Artacho and J. M. Soler, 
    Phys.~Rev.~B \textbf{53}, R10441 (1996); 
    J.M. Soler, E. Artacho, J.D. Gale, A. Garc{\'{\i}}a, J. Junquera,
    P. Ordej{\'o}n, and D. S{\'a}nchez-Portal, 
    J. Phys.: Condens. Matt. \textbf{14}, 2745 (2002). \\
6  & O. Pag\`{e}s, M. Ajjoun, T. Tite, D. Bormann, E. Tourni\'{e}
     and K.C. Rustagi,
     Phys. Rev. B \textbf{70}, 155319 (2004). \\
7 & O. Pag\`{e}s, A.~V. Postnikov, A. Chafi, D. Bormann, P. Simon,
    F. Firszt, W. Paszkowicz and E. Tourni\'e, 
    cond-mat/0610682. \\
\end{tabular}
}
\end{document}